\documentclass[usegraphicx,useAMS,usenatbib]{mn2e}
\usepackage{times}
\usepackage{amssymb}
\usepackage{amsmath}
\usepackage{graphicx}
\usepackage{euscript}
\usepackage{rotating}
\usepackage{epsfig}

\def\hz{{\rm\thinspace Hz}}

\def\ghz{{\rm\thinspace GHz}}

\def\kmpspmpc{\hbox{$\rm\thinspace km~s^{-1}~Mpc^{-1}$}}

\def\mev{{\rm\thinspace MeV}}
\def\G{{\rm\thinspace G}} 

\def\mas{{\rm\thinspace mas}}

\begin{document}

\newcommand{\Mpc}{\rm\thinspace Mpc}
\newcommand{\kpc}{\rm\thinspace kpc}
\newcommand{\pc}{\rm\thinspace pc}
\newcommand{\km}{\rm\thinspace km}
\newcommand{\m}{\rm\thinspace m}
\newcommand{\cm}{\rm\thinspace cm}
\newcommand{\cmps}{\hbox{$\cm\s^{-1}\,$}}
\newcommand{\cmpssq}{\hbox{$\cm\s^{-2}\,$}}
\newcommand{\cmsq}{\hbox{$\cm^2\,$}}
\newcommand{\cmcu}{\hbox{$\cm^3\,$}}
\newcommand{\pcmcu}{\hbox{$\cm^{-3}\,$}}
\newcommand{\pcmcuK}{\hbox{$\cm^{-3}\K\,$}}

\newcommand{\yr}{\rm\thinspace yr}
\newcommand{\gyr}{\rm\thinspace Gyr}
\newcommand{\s}{\rm\thinspace s}
\newcommand{\ks}{\rm\thinspace ks}

\newcommand{\GHz}{\rm\thinspace GHz}
\newcommand{\MHz}{\rm\thinspace MHz}
\newcommand{\Hz}{\rm\thinspace Hz}

\newcommand{\K}{\rm\thinspace K}

\newcommand{\Kpcmc}{\hbox{$\K\cm^{-3}\,$}}

\newcommand{\g}{\rm\thinspace g}
\newcommand{\gpcm}{\hbox{$\g\cm^{-3}\,$}}
\newcommand{\gpcmps}{\hbox{$\g\cm^{-3}\s^{-1}\,$}}
\newcommand{\gps}{\hbox{$\g\s^{-1}\,$}}
\newcommand{\Msun}{\hbox{$\rm\thinspace M_{\odot}$}}
\newcommand{\Msunpc}{\hbox{$\Msun\pc^{-3}\,$}}
\newcommand{\Msunpkpc}{\hbox{$\Msun\kpc^{-1}\,$}}
\newcommand{\Msunppc}{\hbox{$\Msun\pc^{-3}\,$}}
\newcommand{\Msunppcpyr}{\hbox{$\Msun\pc^{-3}\yr^{-1}\,$}}
\newcommand{\Msunpyr}{\hbox{$\Msun\yr^{-1}\,$}}

\newcommand{\MeV}{\rm\thinspace MeV}
\newcommand{\keV}{\rm\thinspace keV}
\newcommand{\eV}{\rm\thinspace eV}
\newcommand{\erg}{\rm\thinspace erg}
\newcommand{\Jy}{\rm Jy}
\newcommand{\ergpcmc}{\hbox{$\erg\cm^{-3}\,$}}
\newcommand{\ergcmcups}{\hbox{$\erg\cm^3\ps\,$}}
\newcommand{\ergpcmps}{\hbox{$\erg\cm^{-3}\s^{-1}\,$}}
\newcommand{\ergpcmsqps}{\hbox{$\erg\cm^{-2}\s^{-1}\,$}}
\newcommand{\ergpcmsqpspA}{\hbox{$\erg\cm^{-2}\s^{-1}$\AA$^{-1}\,$}}
\newcommand{\ergpcmsqpspsr}{\hbox{$\erg\cm^{-2}\s^{-1}\sr^{-1}\,$}}
\newcommand{\ergpcmcups}{\hbox{$\erg\cm^{-3}\s^{-1}\,$}}
\newcommand{\ergps}{\hbox{$\erg\s^{-1}\,$}}
\newcommand{\ergpspmp}{\hbox{$\erg\s^{-1}\Mpc^{-3}\,$}}
\newcommand{\keVpcmsqpspsr}{\hbox{$\keV\cm^{-2}\s^{-1}\sr^{-1}\,$}}

\newcommand{\dyn}{\rm\thinspace dyn}
\newcommand{\dynpcmsq}{\hbox{$\dyn\cm^{-2}\,$}}

\newcommand{\kmps}{\hbox{$\km\s^{-1}\,$}}
\newcommand{\kmpspmp}{\hbox{$\km\s^{-1}\Mpc{-1}\,$}}
\newcommand{\kmpspMpc}{\hbox{$\kmps\Mpc^{-1}$}}

\newcommand{\Lsun}{\hbox{$\rm\thinspace L_{\odot}$}}
\newcommand{\Lsunppc}{\hbox{$\Lsun\pc^{-3}\,$}}

\newcommand{\Zsun}{\hbox{$\rm\thinspace Z_{\odot}$}}
\newcommand{\gauss}{\rm\thinspace gauss}
\newcommand{\arcsecond}{\rm\thinspace arcsec}
\newcommand{\chisq}{\hbox{$\chi^2$}}
\newcommand{\delchi}{\hbox{$\Delta\chi$}}
\newcommand{\ph}{\rm\thinspace ph}
\newcommand{\sr}{\rm\thinspace sr}

\newcommand{\pccm}{\hbox{$\cm^{-3}\,$}}
\newcommand{\psqcm}{\hbox{$\cm^{-2}\,$}}
\newcommand{\pcmsq}{\hbox{$\cm^{-2}\,$}}
\newcommand{\pmpc}{\hbox{$\Mpc^{-1}\,$}}
\newcommand{\pmpccu}{\hbox{$\Mpc^{-3}\,$}}
\newcommand{\ps}{\hbox{$\s^{-1}\,$}}
\newcommand{\pHz}{\hbox{$\Hz^{-1}\,$}}
\newcommand{\pcmK}{\hbox{$\cm^{-3}\K$}}
\newcommand{\phpcmsqps}{\hbox{$\ph\cm^{-2}\s^{-1}\,$}}
\newcommand{\psr}{\hbox{$\sr^{-1}\,$}}
\newcommand{\pspsqas}{\hbox{$\s^{-1}\,\arcsecond^{-2}\,$}}

\newcommand{\ergpspcmpK}{\hbox{$\erg\s^{-1}\cm^{-1}\K^{-1}\,$}}

\title{Using Radio Bubbles to Constrain the Matter
  Content of AGN Jets}\author[Dunn,
Fabian \& Celotti] {\parbox[]{6.in} {R.J.H. Dunn$^1$\thanks{E-mail:
rjhd2@ast.cam.ac.uk}, A.C. Fabian$^1$ and A. Celotti$^2$\\
\footnotesize $^1$Institute of Astronomy, Madingley Road, Cambridge
CB3 0HA\\ $^2$International School for Advanced Study, via Beirut
2-4, 34014 Trieste, Italy }} \maketitle

\begin{abstract}
We revisit a method to obtain upper limits on the jet matter content
combining synchrotron self-absorption constraints and the large scale
bubble energy.  We use both X-ray observations, which give limits on the jet power
from the energies and timescales of bubbles found in clusters of
galaxies, and radio observations, which give limits on the magnetic
field in the jets.  Combining the two imposes constraints on the
particle number density, and hence the jet content.  Out of a sample
of clusters which have clear radio bubbles, there are only two which
have sufficient resolution in the radio images to give significant constraints, under the assumption that the jets are fairly steady.  The
results for M87 and Perseus indicate that the radio emitting region of the jet is
electron-positron dominated, assuming that the minimum of the electron energy
distribution, $\gamma_{\rm min}\sim 1$.
\end{abstract}


\section{Introduction}

The radio synchrotron emission from jets associated with active
galactic nuclei (AGN) indicate that they must contain relativistic
leptons and magnetic fields.  The matter content of these jets,
however, is still a mystery (e.g. \citealt{Homan05}).  Assuming that
these jets are electrically neutral, are they electron-positron
(``light'') or electron-proton (``heavy'')?  This paper aims to
determine whether the positively charged particles are positrons or
protons, in the assumption that the magnetic field is not
energetically dominant.

Previous attempts to determine the matter content of the jets have
used Synchrotron Self-Absorption (SSA) arguments
(e.g. \citealt{CelottiACFjets,ReynoldsACFjets}).  Measurements of the
core radio flux and the size of the emitting region lead to
constraints on the number density of the relativistic particles
responsible for the emission, $n$, and on the magnetic field present
in the jet, $B$.  The effect of any protons in the jet cannot be
determined from radiative information.  As a result the large scale
properties of the source are required to place independent estimates
on the total power.  In the past this has been difficult because the
total energy in the radio lobes has not been known; equipartition
between the energy in the magnetic field and the radio-emitting
particles has been assumed in order to be able to estimate the energy
present \citep{Burbidge}.

With the advent of the {\it Chandra} and {\it XMM-Newton} X-ray
observatories the interaction of radio sources in the centres of
clusters of galaxies with the surrounding Intra-Cluster Medium (ICM)
has been observed in great detail.  One of the first observed was the
Perseus Cluster, where decrements in the X-ray emission from the
thermal ICM were first seen by {\it ROSAT} \citep{Bohringer}.  These
have been interpreted as ``bubbles'' blown in the ICM by the central
radio source.  Strong support for this interpretation came with {\it
Chandra} imaging and spectral information
\citep{ACF_complex_PER00}. Subsequently many other examples of such
bubbles have been found (e.g. Hydra A, \citealp{McNamaraHydra00};
A2052, \citealp{Blanton01}; A2199, \citealp{JohnstoneA2199};
Centaurus, \citealp{SandersCent02}).

As these radio lobes are embedded in the ICM, their energy content can
be more accurately determined using arguments of pressure balance
between the thermal ICM and the relativistic plasma within the
bubbles, as well as bubble dynamics.  As a result, the connection
between the kinetic luminosity of the jet and the energy in the bubble
can be calculated. The potential of this approach is that it does not
require assumptions on equipartition for the estimate of the jet
power.

In this work we combine the information from X-ray and VLBI
observations of radio sources embedded in clusters of galaxies, where
clear bubbles are visible in the X-ray images.  We apply the
calculations performed for the jet in M87 by \citet{ReynoldsACFjets}
to calculate the matter content of the jet present in this and other
sources.  In essence the method determines the lepton content of the
base of the jet from synchrotron emission and absorption, and the
total particle energy content from power required to create the
observed bubbles.  The comparison between the two, sets constraints on
the particle content.

In Section \ref{SynSelAbs} we outline the models and assumptions used
in the calculations.  The source parameters and the results are
presented in Sections \ref{sourceparam} and \ref{results}.  The
assumed lepton energy distribution is discussed in Section
\ref{gamma_min}, and further complications of this method and
comparisons with other methods are discussed in Section
\ref{discussion}.  The implications of this work on the particle
content of radio lobes is investigated in Section
\ref{radio_particle}.  We use \mbox{$H_0= 70 \kmpspmpc$} throughout.

\section{The Model}\label{SynSelAbs}

This section briefly summarises the calculations presented in
\citet{ReynoldsACFjets} (following \citealt{Marscher87Conf}) in order
to demonstrate the method and define the parameters.  

\subsection{Limits on the Magnetic Field from Synchrotron Self-Absorption}

With the
assumption that the relativistic leptons have a distribution of
Lorentz factors \mbox{$N(\gamma)=N_0\gamma^{\rm
-p}=N_0\gamma^{-(2\alpha+1)}$}, between $\gamma_{\rm min}$ and
$\gamma_{\rm max}$, the spectrum in the optically thin regime is
$S_{\nu}\propto\nu^{-\alpha}$.  The corresponding relativistic lepton
number density, $n$, is given by
\begin{eqnarray}
\centering
n &=&  \int^{\gamma_{\rm max}}_{\gamma_{\rm min}}N(\gamma)d\gamma
\nonumber \\
2\alpha n&=&-N_0 \left[\gamma^{-2\alpha} \right] ^{\gamma_{\rm max}}
_{\gamma_{\rm min}}.\nonumber
\end{eqnarray}

\noindent Therefore, for $\gamma_{\rm max}>> \gamma_{\rm min}$ and $\alpha>0.5$, the lepton number density is
\begin{equation}
 n=\frac{N_0}{2\alpha}\gamma_{\rm min}^{-2\alpha}.\label{n_and_N_0}
\end{equation}
Initially we assume $\gamma_{\rm min}=1$ for the following calculations.  For
a discussion of the effect of varying $\gamma_{\rm min}$ see Section
\ref{gamma_min}.

The synchrotron flux in the optically thick (self-absorbed) region is
independent of the particle density, resulting in an estimate on
the magnetic field of
\begin{equation}
B\lesssim 10^{-5}b(\alpha)\theta^4_{\rm d}\nu_{\rm m}^5S_{\rm
    m}^{-2}\Bigl(\frac{\delta}{1+{\rm z}}\Bigr) {\rm G},\label{Blimit}
\end{equation}
where the $\nu_{\rm m}$ and $S_{\rm m}$ are the frequency (in $\ghz$)
and flux density (in Jy) at spectral turnover.  $\theta_{\rm
d}$ is the angular diameter of the source in milliarcseconds (mas).
As the angular diameter may be an upper limit (the core may not be
fully resolved) this causes the estimate on $B$ to be an upper limit.
In some cases the source is elliptical, and in these cases the average
\mbox{$\theta_{\rm d}=\sqrt{\theta_{\rm a}\theta_{\rm b}}$}, where
$\theta_{\rm a}$ and $\theta_{\rm b}$ are the corresponding angular
diameters of the ellipse.  $\delta$ is the relativistic Doppler
factor, defined as \mbox{$\delta=1/\Gamma(1-\beta\cos\phi)$} where
$\Gamma$ and $\beta$ are the Lorentz factor and $v/c$ for the bulk
motion of the jet respectively and $\phi$ is the angle between the
line of sight and the jet axis.  The function,
$b(\alpha)$, is tabulated in \citet{Marscher87Conf} and we have
interpolated between the values given where appropriate. 

The observation of compact radio sources with flat spectra,
\mbox{$\alpha_{\rm obs}\sim 0$}, has been interpreted as the
superposition of different SSA components each peaking at different
frequencies \citep{Blandford_79_jets}. Hence the observations at
a given $\nu$ basically measure the flux density and the size of the
component of the jet which is becoming self absorbed at $\nu$ (in the
observer's frame).

\subsection{Limits on the Magnetic Field and Number Density}

The jet becomes self-absorbed at an observed frequency,
\mbox{$\nu_{\rm m}$}, where \mbox{$\tau_{\rm syn} (\nu_{\rm m}, r) =
\kappa(\nu_{\rm m})X = 1$}; where $\kappa(\nu)$ is the 
synchrotron absorption coefficient, $X$ is the path length of the
line of sight through the jet and $r$ the jet cross-section radius.
From \citet{ReynoldsACFjets} and references therein
\begin{equation}
\kappa(\nu_{\rm m})=\frac{3^{(\alpha+1)} \sqrt{\pi} g(\alpha) e^2
      N_0}{8m_{\rm e}c}\nu_{\rm B}^{(3/2+\alpha)}\nu_{\rm
      m}^{-(5/2+\alpha)}\delta^{(5/2+\alpha)} ,\label{kappa}
\end{equation}
where $\nu_{\rm B}$ is the cyclotron frequency and $g(\alpha)$ is the
product of gamma functions (of order unity for the considered range of
$\alpha$).

This expression for $\kappa(\nu)$ is valid for \mbox{$\nu\gg\nu_{\rm
min}\sim\gamma^2_{\rm min}\nu_{\rm B}$}, where $\nu_{\rm min}$ is the
low-energy cut-off in the spectrum corresponding to the low-energy
cut-off in the lepton energy distribution.  In this case the
self-absorption only depends on the normalisation of the relativistic
lepton distribution and the magnetic field.

The path length is $X=2r/\delta$,
using the relativistic transformations for a cylindrical geometry.
Combining this with Equations \ref{n_and_N_0} and \ref{kappa}, places
a lower limit in the $B-n$ plane for radiation at frequency $\nu_{\rm m}$ to
be self-absorbed in the source,

\begin{equation}
nB^{(\frac{3}{2}+\alpha)}\gtrsim \frac{2 \delta}{3^{(\alpha
    +1)}\sqrt{\pi} g(\alpha) \alpha \gamma_{\rm min}^{2\alpha} e
    r}\Big(\frac{m_{\rm e} c \nu_{\rm m}}{e\delta}\Big)^{(\frac{5}{2}
    + \alpha)}.\label{nBplane}
\end{equation}

\subsection{Kinetic Luminosity}\label{KinLum}

The kinetic luminosity of the jet depends on the type of particles
present, as well as their energy and number density.  In the
assumption that all of the energy contained within the jet results in
the creation and expansion of the radio bubbles observed within
clusters, then the power required to create these bubbles is an
estimator of the (average) kinetic luminosity of the jet,
i.e. \mbox{$L_{\rm K}=E_{\rm bubble}/t_{\rm bubble}$}, where $t_{\rm
  bubble}$ is the creation time of the bubble.  The simplest
estimate on the energy required is that $E_{\rm bubble}=pV$, where $V$
is the volume of the bubble, and $p$ is the pressure of the
surrounding intra-cluster gas.  Taking into account any internal
energy of the bubble results in
\[
E_{\rm bubble}=\frac{\gamma_{\rm R}}{\gamma_{\rm R}-1}pV,
\]
where $\gamma_{\rm R}$ is the ratio of specific heat capacities, which
for a relativistic gas is $4/3$, resulting in $E_{\rm bubble}=4pV$.
Whether the energy contained within the bubble is $pV$, $4pV$ or some
other multiple of $pV$ is currently uncertain.  For example,
investigating the weak shock surrounding the bubbles in the Perseus
cluster, \citet{ACF_Per_Mega_06} find that the energy of the
post-shock gas is around $2pV$.  We use $E_{\rm bubble}=pV$, although
using $4pV$ does not change the results significantly.

In our analysis we assume that the radio bubbles are in pressure
equilibrium with their surroundings, whereas \citet{ReynoldsACFjets}
assumed that they were over-pressured by a factor of $\sim3$.  Using
their source parameters we recover the limits they place on the matter
content of the M87 jet (see their Fig. 1).

There are a number of estimates on the bubble timescales.  The most
appropriate one for these young bubbles which are (presumably) still
being inflated by their jet is the sound speed timescale, $t_{\rm bubble}=2R_{\rm
bubble}/c_{\rm s}$, where $R_{\rm bubble}$ is the bubble radius and
$c_{\rm s}$ is the local sound speed, following \citet{DunnFabian04}.  There are no indications for strong shocks
surrounding the bubbles, which implies that the bubble edges are
travelling at less than the local sound speed.   It is
possible, however, that the bubbles do not grow smoothly, but in fits
and starts \citep{Fabian_05_visc_cond}, and as such this timescale is
not a good estimate for the age of the bubble.  For further discussion
on timescales relevant to the evolution of bubbles in clusters see
\citet{Churazov00,DunnFabian04,DunnFabian05}.

As mentioned, we assume that the jet is particle dominated
(i.e. we neglect the magnetic field contribution) and for simplicity,
following other work (e.g. \citealt{Sikora00}), assume that the
protons are ``cold''. Thus the jet kinetic luminosity, including the
advected energy, is
\begin{equation}
L_{\rm K}\approx\Gamma^2 \beta \pi r(Z)^2 n m_{\rm e}c^3
\left[\frac{4}{3}(\langle\gamma\rangle-1)+\frac{\Gamma-1}{\Gamma}
(1+k_{\rm a})\right]
\label{LK}
\end{equation}
\noindent 
where $k_{\rm a}$ takes into account the effect of
hadrons on the rest-mass energy (adapted from \citealt{Schwartz06}).
For electron-positron jets $k_{\rm a}=0$.  In electron-proton jets we
assume that there is one proton per electron, so $k_{\rm a}=m_{\rm
p}/m_{\rm e}$.  The internal energy of the leptons is expressed as
$(\langle\gamma\rangle-1)n m_{\rm e}c^2$ $\sim \langle\gamma\rangle n
m_{\rm e}c^2$ for \mbox{$\langle\gamma\rangle >>1$}.

If $\alpha=0.5$ then the average particle Lorentz factor is
\[
\langle\gamma\rangle=\gamma_{\rm min}{\rm ln}\!\left(\frac{\gamma_{\rm
max}}{\gamma_{\rm min}}\right).
\]
For a more general case, where $\gamma_{\rm max}>> \gamma_{\rm min}$,
\[
\langle\gamma\rangle=\frac{2\alpha\gamma_{\rm
    min}^{2\alpha}}{1-2\alpha}\left(\frac{1}{\gamma_{\rm
    max}^{(2\alpha-1)}}-\frac{1}{\gamma_{\rm
    min}^{(2\alpha-1)}}\right) \approx \frac{2 \alpha}{2\alpha-1}
    \gamma_{\rm min}.
\]

Using the bubble power and Equation \ref{LK} estimates on the number
density can be calculated for the two different jet compositions.
These expected number densities are compared to the limits imposed on
the $B-n$ plane by Equations \ref{Blimit} and \ref{nBplane}.

\section{Source Parameters}\label{sourceparam}

Here, as a result of the range of spectral indices measured, we assume
that each SSA component has the same particle energy distribution,
with $p=2.4$ and so the same (optically thin) spectral index
$\alpha=0.7$, as this falls in the middle of the range expected for
optically thin spectra ($0.5\lesssim\alpha\lesssim1.0$).

The core fluxes presented in this work are those which place some
useful limit on the matter content of the jets.  There are many other,
mainly older, measurements which result in lower limits on the number
densities inferred which are too low to allow any constraints to
be placed on the matter content.

\begin{table}
\centering
\caption{\scshape \label{Cluster}: Cluster Values and Notes}
\begin{tabular}{lcccc}
\hline
\hline
Bubble&$\phi$&$E_{\rm bubble}$&$t_{\rm bubble}$\\
&Degrees&$10^{57}\erg$&$10^7 \yr$\\
\hline
\hline
\multicolumn{5}{c}{M87, $z=0.004$}\\
\hline
Jet        &$15.0\pm5.0 $   &$0.28  \pm0.14$  &$0.59  \pm0.03$\\   
Counter-Jet&$15.0\pm5.0 $   &$0.39  \pm0.20$  &$0.48  \pm0.02$\\   
\hline
Average&&$0.33\pm0.12$&$0.53\pm0.02$&\\
\multicolumn{5}{l}{\hspace{1cm}$L_{\rm K}=2.0\times10^{42}\ergps$}\\
\hline
\hline
\multicolumn{5}{c}{Perseus, $z=0.018$}\\
\hline
Northern-Inner&$40.0	\pm10.0$       &$4.97  \pm2.20$	  &$0.98  \pm0.18 $\\ 
Southern-Inner&$40.0	\pm10.0$       &$4.32  \pm1.98$	  &$0.93  \pm0.18 $\\
\hline
Average&&$4.64\pm1.48$&$0.95\pm0.13$&\\
\multicolumn{5}{l}{\hspace{1cm}$L_{\rm K}=1.5\times10^{43}\ergps$}\\
\hline
\end{tabular}
\begin{quote}
{\scshape References:} The energies and timescales for M87 and Perseus
are calculated from the bubble radii presented in \citet{Allen_Bondi_06} and \citet{Precession} respectively.
The angles to the line of sight are from \citet{Biretta_99} and
\citet{Walker94}.  \citet{ReynoldsACFjets} used $L_{\rm
K}=10^{43}\ergps$.
\end{quote}
\end{table}

\begin{table*}
\caption{\scshape \label{Components}: Core fluxes and jet properties}
\begin{tabular}{lccccccc}
\hline
\hline
Name & Date & Flux Density (Jy) & \multicolumn{2}{c}{Size
  (mas); $\theta_{\rm a}$, $\theta_{\rm b}$}&$\Gamma^{\rm a}$&
$\nu(\ghz)$&Reference \\
\hline
\multicolumn{8}{c}{M87}\\
\hline
Reynolds$^{\rm b}$      & 0972/0373  &$1.00\pm  0.10$    &$0.70  \pm  0.07   $  &$0.70   \pm0.07   $ &$3\pm1$  &5.0 &  1,2,3,4\\	
Kovalev-$S_{\rm Core}$	& 050203     &$1.39\pm  0.07$    &$0.41  \pm  0.04   $  &$0.27   \pm0.03   $ &$10\pm5$ &15.0&  5,6 \\
Kovalev-$S_{\rm Unres}^{\rm c}$ & 050203     &$0.73\pm  0.04$    &$0.41  \pm  0.04   $  &$0.27   \pm0.03   $ &$10\pm5$ &15.0&  5,6 \\
Lister05-$I_{\rm c}$	& 050203     &$0.98\pm  0.05$    &$0.37  \pm  0.02   $  &$0.14   \pm0.01   $ &$10\pm5$ &15.0&  6,7 \\
Scott-$S^{\rm c}$ 	& 201297     &$0.18\pm  0.03$    &$0.20  \pm  0.02   $  &$0.20   \pm0.02   $ &$10\pm5$ &5.0 &  6,8\\
\hline
\multicolumn{8}{c}{Perseus}\\
\hline
Kovalev-$S_{\rm Core}$	& 010303 &$3.63 \pm   0.18 $    &$0.66  \pm  0.07   $ & $0.21	\pm 0.02$   & $1.10\pm0.05$ &15.0 &  5,9\\
\hline
\hline	

\end{tabular}
\begin{quote}
\noindent
{\scshape Notes:} All fluxes and sizes are model fits to the
emission.  The beam for Lister05 is $1.23\times0.55$~mas.  Date is in DDMMYY or MMYY/MMYY format. $^{\rm a}$ The $\Gamma$ is for the
bulk motion of the jet. $^{\rm b}$ $\alpha=0.5$ for this core flux.
$^{\rm c}$ The angular sizes are an upper limit for these
measurements.  
{\scshape References:}  
1. \citet{Pauliny-Toth81};
2. \citet{Biretta_93_bk};
3. \citet{Biretta_M87_1995};
4. \citet{ReynoldsACFjets};
5. \citet{Kovalev05};
6. \citet{Biretta_99}
7. \citet{Lister05};
8. \citet{Scott04};
9. \citet{Walker94}.
\end{quote}
\end{table*}

\subsection{M87}\label{m87}

We include the VLBI and $L_{\rm K}$ parameters adopted by
\citet{ReynoldsACFjets} (using $\alpha=0.5$) to allow a
comparison to be made between their values and the values based on
updated observations. The jet parameters used \citet{ReynoldsACFjets}
were $\Gamma\gtrsim3$ and $\phi\sim 30^\circ$, based on the
observations in \citet{Biretta91_M87} and \citet{Biretta_M87_1995}.

The apparent jet speeds measured by \citet{Biretta_M87_1995} ranged
from $2c$ close to the core to $\sim0.5c$ for more distant components.
Recent observations report a range of apparent speeds;
\citet{Kellermann04} measure $0.04\pm0.02c$ for a component $6\mas$
from the core; \citet{Ly_AAS_04} place lower limits of $0.25-0.40c$ on
the jet motion and \citet{Dodson_05} find little evidence for any
motion of the components close ($<150\mas$) to the core. A series of
HST images resulted in measurements of apparent speeds of $2.6-6c$ for
components between $0.87''$ and $6''$ from the core
\citep{Biretta_99}, with the innermost resolved component at $160\mas$
from the core having an apparent speed of $0.63\pm0.27c$.

Clearly the superluminal velocities measured might not be associated
with the bulk flow of the jet, but correspond to pattern speeds
(e.g. shocks). However, if the observed speeds do arise from the
plasma bulk motion, then the above measures place strong constraints
on its velocity and the angle of the jet to the line of sight.
\citet{Biretta_M87_1995} found that $\phi<43^\circ$.  Assuming that
the apparent speeds are $\sim 6c$ then $\phi<19^{\circ}$
\citep{Biretta_99} (further allowed combinations of $\Gamma$ and
$\phi$ are shown in their Table 3).  As a result we use
$\Gamma=10.0\pm5.0$ as this encompasses a large proportion of the
range of observed superluminal speeds, along with a corresponding
$\phi=15\pm5^\circ$.

\citet{Dodson_05} measure a spectral index of $-1.1\pm0.4$ between 1.6
and 4.8\ghz, and $-0.6\pm0.4$ between 4.8 and 15\ghz.  Their
calculations of the spectral turnover gives $\sim 20\ghz$.  The core
flux densities used here have been measured at $5$ and $15\ghz$.
These are close to the spectral turnover frequency as estimated by
\citet{Dodson_05} and are thus presumably in the self-absorbed region
of the spectrum, where Equation \ref{kappa} is applicable.

\subsection{Perseus}\label{perseus}

Studies of the core of 3C~84 have also resulted in a range of measured
apparent motions.  \citet{Kellermann04} using $15\ghz$ VLBA
observations find $\beta=0.2\pm0.1$.  \citet{Walker94} find that the
jet speed and orientation angle are not tightly constrained,
$\beta=0.3-0.5$ and $\phi=30^\circ-55^\circ$.  As in M87, some
components appear to be much slower, $\beta\sim0.05$ \citep{Dhawan98},
and others are faster, $\beta=0.7\pm0.1$ \citep{Marr89_Per}.
\citet{Krichbaum92_Per} estimate an angle to the line of sight
of $\sim10^\circ$ and a corresponding $\beta>0.68$.  We use the values
presented in \citet{Walker94} as these fall towards the middle of the
observed range ($\beta=0.3-0.5$ corresponds to $\Gamma=1.10\pm0.05$).

The core of 3C84 in Perseus has a complicated spectral index structure
with an inverted spectrum at lower frequencies
($\alpha_{1.5-5.0\ghz}\sim 1$, \citealt{Unwin82_Per,ODea84}),
flattening towards higher frequencies $\alpha_{10-100\ghz}\sim 0$
(\citealt{ODea84}). The spectra in \citet{Readhead83_Per} also show a
decrease of flux density toward lower frequencies, though there are no
measurements at less than $10\ghz$.  The spectral index varies across
the core and so an exact value is difficult to obtain
(e.g. \citealt{Unwin82_Per,Readhead83_Per,Vermeulen94_3C84,
Krichbaum92_Per}). \citet{Silver98_Per} indicate that the spectrum may
indeed be flat in the range $1-10\ghz$ (see their Fig. 5).  Our core
flux for Perseus is at $15\ghz$ and so it is reasonable that Equation
\ref{kappa} is applicable.

\section{Results}\label{results}

\begin{figure}
\centering
\includegraphics[width=0.45\textwidth]{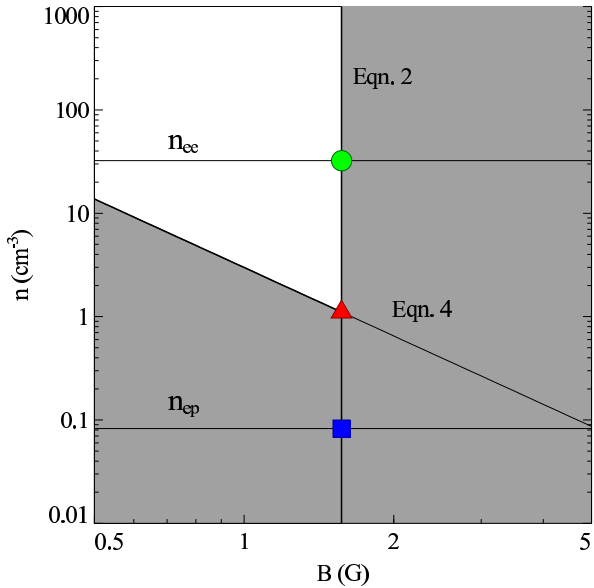}
\caption {\label{schematic}\small{Constraints on the $B-n$ plane from
  Equations \ref{Blimit} and \ref{nBplane}.  The expected values for
  the number density for light and heavy jets from the estimated
  $L_{\rm K}$ are shown and correspond to the circle and
  square respectively.  The ``intersection point'' (triangle)
  is the number density required from radio observations.
  This schematic corresponds to Kovalev-$S_{\rm Unres}$ from M87, and
  for clarity the uncertainties are not shown.}}
\end{figure}

Fig. \ref{schematic} shows, for a representative case, the constraints
on the matter content of jets placed by Equations \ref{Blimit} and
\ref{nBplane} in the $B-n$ plane.  The lines labelled ``Eqn. 2'' and
``Eqn. 4'' exclude regions of the $B-n$ plane.  The ``intersection point'' (triangle) is 
thus the minimum number density possible in the jet from observations, 
being the combination of the two above equations. The expected number
densities inferred from the estimated $L_{\rm K}$ for pure
electron-positron (the circle and line labelled ``$\rm
{n_{ee}}$'') and electron-proton (the square and line labelled
``$\rm {n_{ep}}$'') jets are shown. In this representative case an
electron-proton jet is excluded, and hence the jet is
electron-positron.  Note that if the estimate on the magnetic field
were to increase then the intersection point would move to the right,
eventually allowing the possibility that the jet is electron-proton.
When this occurs, no constraints can be placed on the matter content
of the jets. 

\begin{figure}
\centering
\includegraphics[width=0.43 \textwidth]{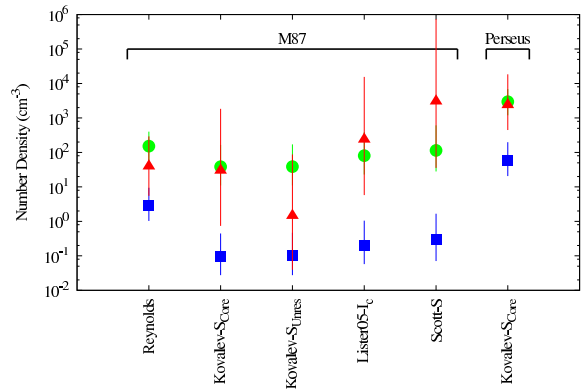}
\caption {\label{results_figure}\small{The relative number densities
      in the jet \emph{Circle} for electron-positron and
      \emph{Square} for electron-proton.  \emph{Triangle}
      makes the intersection of the two limits in the $B-n$ plane and
      represents a lower limit.}}
\end{figure}

The results are shown in Fig. \ref{results_figure}, for both M87 and
Perseus.  In order to compute $L_{\rm K}$ we average the energies and
timescales of the bubbles within each cluster, as the differences in
the results between the different bubbles are very small.

Fig. \ref{results_figure} reports the intersection points
corresponding to the $n-B$ plane for each useful set of data.  
In order to check the robustness of the conclusions, we also
explicitly estimated errors on the inferred values. In particular, 
the
reported uncertainties are estimated using a Monte-Carlo algorithm,
assuming the input uncertainties have a Gaussian distribution which
take into account range of values measured for each parameter. The
source of uncertainties which have been accounted for are those parameters which have an uncertainty quoted in
Tables \ref{Cluster} and \ref{Components}.  If no uncertainty has been given in the
literature, we have adopted a conservative value of 10 per cent ($\alpha$ is assumed
to have no error, but the effect of different values is discussed in
Section \ref{param_effects}).  The error-bars shown on
Fig. \ref{results_figure} are $1-\sigma$ errors derived from this
algorithm.  These uncertainties are only applicable to this model for
the core radio emission.

Although the different datasets imply different values of densities,
globally they provide a similar result; that the relativistic jets in
these two sources are likely to be electron-positron.  This analysis indicates that for M87, including the core flux used in
\citet{ReynoldsACFjets}, the jet is electron-positron.  There is,
however, only one measurement for the Perseus cluster (from
\citealt{Kovalev05}) which places clear limits on the matter content
of the jet in 3C~84.

As a result of the large uncertainties in the measurements of the jet
angles and bulk velocities and the disparity between the values for
M87 and Perseus we check if the conclusions change substantially using
different $\Gamma$ and $\phi$.  We recalculate the results for M87
with $\Gamma=1.1\pm0.05$ and $\phi=35\pm5^\circ$ and for Perseus with
$\Gamma=10\pm5$ and $\phi=10\pm5^\circ$.

Relative to the number densities required by the estimated
$L_{\rm K}$ (which themselves rise by a factor $\times 10^2-10^3$)
the intersection points (red triangles) for M87 fall.  The results
still favour light jets, though are less definite.  In Perseus, as the
jet angle and velocity were altered in the other sense, the number
densities corresponding to $L_{\rm K}$ decrease and the limits
become more consistent with light jets.

Thus, even though we have chosen values for $\Gamma$ and $\phi$
which are drastically different from the average of the measured
values the results still indicate that both jets are light, showing
that the conclusions do not depend critically on these values.

Other clusters were also analysed, including A2199 (3C~338) and
Hydra A, but the radio observations of the core did not have
sufficient resolution (the model is very dependent on the angular size
of the core).  The resultant magnetic field estimates were
therefore high, giving very low estimates on the number densities in
the jet.  Therefore no useful constraints could be placed on the
matter content.

\section{$\gamma_{\rm\MakeLowercase{ min}}$ Limits}\label{gamma_min}

We have so far assumed that the low-energy cut-off in the energy
distribution of leptons occurs at $\gamma_{\rm min}\sim 1$.
\citet{CelottiACFjets}, \citet{Ghisellini1992} and
\citet{ReynoldsACFjets}, however, find that a low-energy cut-off at
$50\mev~(\gamma_{\rm
min}\approx100)$ in the lepton energy spectrum is required for both heavy and
light jets, consistent with polarisation measurements.

As $\gamma_{\rm min}$ rises then the number densities will fall as
fewer particles will carry the same energy (the average energy per
particle will rise).  In the assumption that the protons remain cold,
one can even reach the situation where the average lepton energy
is close to the proton energy (i.e. $\langle\gamma\rangle \simeq
\frac{m_{\rm p}}{m_{\rm e}}$). It is clear that when approaching this value,
the expected lepton number densities for the two jet types
differ by a smaller and smaller factor and so, given the large
uncertainties of the current measurements, by increasing $\gamma_{\rm
min}$ the possibility of drawing any conclusions on the
matter content becomes increasingly small.

\subsection{Synchrotron and Synchrotron Self-Compton constraints}

Upper limits on $\gamma_{\rm min}$ in the core, however, can be set.

A first constraint on maximum values of $\gamma_{\rm min}$ is set by
the very same self-absorption model adopted above, as, for it to be
self-consistent, the lepton distribution should extend at least to
energies of the leptons responsible for the emission at $\nu_{\rm m}$.
In other words, $\gamma_{\rm min}<(3 \nu_{\rm m}/4 \nu_{\rm
B})^{0.5}$. 

Further upper limits can be also obtained from
synchrotron Self-Compton (SSC) constraints.  Namely, the X-ray
emission expected from the self-absorbing region via SSC, $S_{\rm X}$
should not exceed the observed X-ray flux, where
\begin{equation}
S_{\rm X}=\frac{2\alpha S_{\rm m} n \sigma_{\rm T} r \gamma_{\rm
    min}^{2\alpha}}{t(\alpha)}\left(\frac{\nu_{\rm m}}{\nu_{\rm
    X}}\right)^\alpha {\rm ln}\left(\frac{\nu_{\rm b}}{\nu_{\rm
    m}}\right)
\end{equation}\label{SSC}
(e.g. \citealt{ReynoldsACFjets,Marscher87Conf}).  Here $\sigma_{\rm
T}$ is the Thomson scattering cross-section, and $t(\alpha)$ is
tabulated in e.g. \citet{Ghisellini1992}.  The break frequency/upper
cut-off in the synchrotron spectrum, $\nu_{\rm b}$, is taken to lie at
$\sim 10^{15}\hz$ \citep{Meisenheimer_M87_96} and $\nu_{\rm X}$ is the
X-ray frequency.  It is of course possible that the X-ray emission is  not entirely
the result of the SSC process.  Some may be synchrotron emission
(e.g. \citealt{Marshall_M87_2002}), in which case $S_{\rm X}$ is an
upper limit to the SSC flux.

\begin{figure}
\centering
\includegraphics[width=0.43 \textwidth]{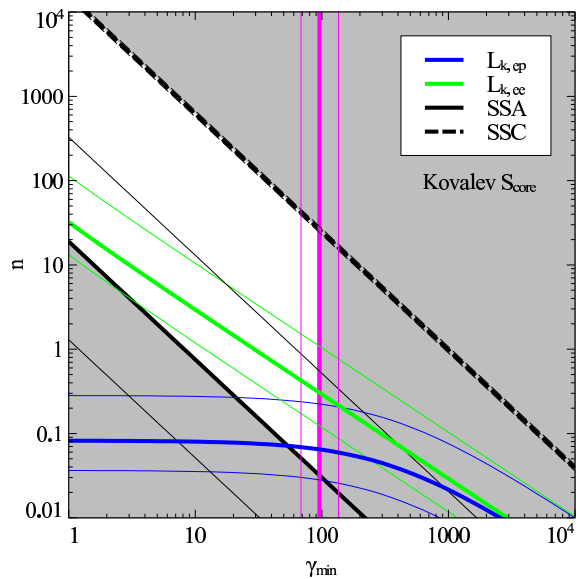}
\caption {\label{gamma_min_fig}\small{$n - \gamma_{\rm min}$ parameter space. The thick lines report the constraints 
imposed by the four conditions: lower limits from SSA (solid black
line); upper limits from SSC (dashed black line); jet kinetic power
from electron-positrons (green line) and electron-proton (blue line);
upper limit from the Larmor frequency $\nu_{\rm B}$ (vertical magenta line). 
The thin lines define the upper and lower uncertainties on each constraint as
derived via the Monte Carlo simulations. The figure illustrates the case for the Kovalev-$S_{\rm Core}$ dataset for M87.}}
\end{figure}

These limits are, as for the SSA, dependent on  $n \gamma_{\rm min}^{2\alpha}$, while the kinetic luminosity 
depends on $n \langle \gamma \rangle$. It is thus necessary  to compare the above constraints in the 
$n- \gamma_{\rm min}$ parameter space for  $\gamma_{\rm min}>1$.
Figure  \ref{gamma_min_fig} reports the results for a representative case (referring to the dataset
Kovalev-$S_{\rm Core}$ of M87), each thick line referring to the upper
limit from SSC (dashed black), the lower limit from SSA (solid black), the 
values corresponding to the inferred $L_{\rm K}$ for pair (green) and
electron-proton (blue) jets, and the upper limit on 
$\gamma_{\rm min}$ from $\nu_{\rm B}$ (magenta).

Taken at face value, the implication is that all the conditions can be satisfied by 
either a pair jet with $1< \gamma_{\rm min}< 100$ or a heavy jet for the limited range 
$50< \gamma_{\rm min}< 100$.  Qualitatively similar inferences can be derived for M87 in the cases of 
Kovalev-$S_{\rm Unres}$ and Reynolds (fully in agreement with the
conclusion reported by \citet{ReynoldsACFjets}), while the Lister-$I_{\rm c}$ is just marginally consistent with 
pair jets with $10< \gamma_{\rm min}< 150$ and Scott-$S$ does not
formally allow a consistent solution.  These last two core flux
measurements are those in which the minimum number density fell above
that allowed for an electron-positron jet in Fig. \ref{results_figure}.  For all cases the SSC upper limits do not give 
interesting constraints. 
On the contrary the Perseus X--ray data (SSC) strongly constrain the possible solutions to light jets with 
$\gamma_{\rm min}\lesssim10$ or heavy jets with $10\lesssim\gamma_{\rm
  min}\lesssim 50$. Overall,  $\gamma_{\rm min}<10$ would imply electron-positron jets. For 
higher values, typically $10<\gamma_{\rm min}<100$ both heavy and
light jets are allowed by the data.  These $\gamma_{\rm
min}$ limits are, on average, not inconsistent with those put forward
by \citet{Celotti02}.  It should be nevertheless stressed the 
(in some cases implausibly) tight range of values of $\gamma_{\rm min}$ were the jets heavy.

The errors on the source parameters, however, limit the possibility of drawing any firm conclusion at this time. 
This can be  seen in Fig. \ref{gamma_min_fig}, where the errors on the above constraints, again derived via the Monte Carlo simulations, 
are reported as thin lines. 
It should also be noticed that the condition imposed by $\nu_{\rm B}$
could be in principle more relaxed if the upper limit on $B$ has been 
overestimated because of the limited resolution of the observations, as   
$\gamma_{\rm min}\propto B^{-1/2} \propto \theta^{-2}$ (Equation \ref{Blimit}).

We conclude that currently this method cannot definitely rule out electron-proton jets in favour of
electron-positron jets, although the former case would require a rather narrow range of values of 
possible $\gamma_{\rm min}$. Better quality data (angular resolution, jet speed, 
jet direction, contribution of SSC emission in the X-ray band) and/or other constraints on 
$\gamma_{\rm min}$ could impose firmer conclusions.

\section{Discussion}\label{discussion}

In the following we will summarise and discuss the caveats of the
above estimates, from both observational and modelling points of view.

At a general level, the conditions on small scales in the jet may not correspond to those
on large scales, from conservation of particles, momentum and energy
alone.  There may be entrainment of ambient material by the jet,
dissipation, turbulence or Poynting flux energy transport.

The jet of M87 has filamentary structure on all scales where it is
well imaged.  There is no obvious distance from the nucleus where the
jet disappears.  If there are any inhomogeneities on small scales in
the jet, then the synchrotron self-absorption estimates could be affected.

\subsection{Variability \& Kinetic Luminosity} 

These calculations assume that the jet is of constant power,
corresponding to that observed now.  This is not necessarily true.  As
the synchrotron flux of the core has been used to determine the matter
content of the jet, any flux variability would change this (both
density and magnetic field), particularly so if the size of the
emitting region appears the same (see Equation \ref{Blimit}). It
would, however, have no immediate effect on the bubble, and hence the
inferred kinetic luminosity would remain the same.

As $B\propto S_{\rm m}^{-2}$ and $nB^{3/2+\alpha}=nB^{2.2}={\rm
const}$ (Equation \ref{nBplane}), then $n\propto S_{\rm m}^{4.4}$ (for
a constant $\theta$ and $\alpha=0.7$).  So a small change in the core
flux changes the minimum possible number density in the jet, even by
orders of magnitude.

To see by how much the core would have to vary for the number
densities in M87 to be consistent with electron-proton jets the core
flux for the Kovalev-$S_{\rm Unres}$ and Scott-$S$ values were
decreased until the intersection point is consistent with the
electron-proton values.  No other values were changed except the
flux. The Kovalev-$S_{\rm Unres}$ core flux only has to change by a
factor of 4 for the number densities to be consistent with
electron-proton jets.  In the case of Scott-$S$ the variability has to
be around a factor of 8.

Indeed, the flux of 3C84 steadily increased between 1960 and 1989
\citep{Marr89_Per,ODea84}, by up to a factor of 4. More recent
observations in the optical and X-ray show that the nuclear flux
varied by less than a factor of 3 between 2000 and 2003
\citep{Perlman_M87_03}. 

This is a major problem with the method utilised in this work. In
principle, there is no temporal link between the current kinetic
luminosity of the jet and that calculated from the bubble. The radio
core and the jet fluxes may vary over the course of decades whereas
the bubbles have timescales of $\sim10^7\yr$.  However recent work by
\citet{Allen_Bondi_06} found a correlation between the Bondi
accretion power and the kinetic luminosity of jets in nine nearby
elliptical galaxies.  The bubble powers are measured over a timescale
of $\sim10^7\yr$ and the inflow through the Bondi radius takes
$\sim10^4-10^5\yr$.  Therefore, the resulting correlation suggests
that these powers might be comparatively steady over $\sim 10^7\yr$. 

Further to the discussion on the bubble powers (Section \ref{KinLum})
is the issue of any energy in the form of cosmic rays and magnetic
fields present at the centres of clusters.  \citet{Sanders05} show
that the pressure of these components could contribute at least 30 per
cent in the region of the Perseus bubbles.  As a result the energy
content of the radio bubbles may be larger than just what
thermodynamic calculations imply, requiring a larger number density of
particles in the jet. There may also be losses as the jet
travels out from the central AGN into the lobe.  The kinetic
luminosity inferred from the bubble expansion would be an under
estimate of the initial kinetic luminosity of the jet.  As such the
expected number densities may be lower limits.  These effects
could cause core fluxes which currently exclude heavy jets to become
consistent with them, but only if these effects are (unexpectedly)
large ($\times\sim100$).

\subsection{Parameter Effects}\label{param_effects}

As was presented in Sections \ref{m87} and \ref{perseus} there is no
clear, exact consensus on the value for all the observational parameters 
which are used in this model.  The true jet velocity and angle to the line of
sight are still uncertain, and the range in the currently inferred
values is large.

The jet velocity affects the expected number densities (Equation
\ref{LK}), $L_{\rm K}\propto n \Gamma^2$ and the magnetic field
estimate (Equation \ref{Blimit}), $B\propto\delta$.  Therefore, if
the true jet velocity were faster, then the
expected number densities required to create the observed bubbles would fall.
If so,  the two core flux measurements for the Perseus jet not
presented in the main results (from \citealt{Kovalev05} and
\citealt{Scott04}) could also imply electron-positron jets (see also
Section \ref{results}).

The angle to the line of sight changes the magnetic field calculated
in the jet, via the Doppler factor $\delta$.  The change (increase) in $\delta$
becomes greater as the angle to the line of sight $\phi$ decreases.  As a result the upper limit
on the magnetic field increases and  the ``intersection
point'' decreases, making the constraints on the matter content
less restrictive.  We have, however, taken into account the large
uncertainties on $\Gamma$ and $\phi$ in our calculations of the errors
associated with the estimated number densities.

The spectral index of the optically thin spectrum, $\alpha$, has
instead very
little effect on the results in the range
$0.5\lesssim\alpha\lesssim1.0$.  As such the conclusions remain
unchanged over a large range in $\alpha$ which is a parameter whose
value we have assumed rather than measured.

\subsection{Dimension of the Self-Absorbed Core}

The angular size of the core has been taken as the diameter of the jet
at the point when the jet becomes self-absorbed; however the core may
not be resolved and so this may be an \emph{upper limit}.  Although we
have used the most recent high resolution data, the actual emission
region might be smaller than the size of the radio telescope beam,
therefore we would have over-estimated the angular size of the source.
As $B\propto \theta^4$ (Equation \ref{Blimit}), the effect of any
difference between the true and observed value would be greatly
amplified in the estimates of $B$ and $n$. Indeed, the sources with
large estimates of magnetic field (e.g. A2199, Hydra A) were also
those furthest away, implying the resolution may not have been high
enough to fully resolve the emitting regions.

Another caveat is that the angular size of the core may also represent
the entire projected length of the self-absorbed jet
\citep{Marscher87Conf}.  However, the most recent observations of M87
have a resolution of 0.1 milliarcsecond \citep{Krichbaum_2004}, which
corresponds to tens of light days.  This means that at least for the
closest objects the core radii are unlikely to be blends of knots of
emission, masking the true core emission.

\subsection{Comparison with other studies}

\citet{Ghisellini1992} used spectral information and VLBI maps from
105 sources along with an SSC model to obtain estimates on the bulk
Doppler factor and emitting particle number density.  Upper limits on
the pair number density were placed from the number expected to be
able to survive annihilation from the central region where they might
be produced, and lower limits from the observed synchrotron emission.
They found that a low-energy cut-off of $\gamma_{\rm
min}\approx100$ in the lepton energy distribution was required for
both light and heavy jets.

Using the same sample, \citet{CelottiACFjets} found that the estimated
kinetic power on parsec scales was similar to that inferred from 
radio lobes, concluding that energy can be efficiently transported over several
orders of magnitude in length scale along the jet.  Combining limits
on the total kinetic power and particle number density flux, their results
favoured heavy jets, also with a low-energy cut-off at around
$\gamma_{\rm min}\approx100$.

From X-ray observations of blazars associated with optically violent
variable quasars, \citet{Sikora00} place constraints on the pair
content of the radio-loud quasar jets.  They exclude both pure light
jets, as these over-predict the soft X-ray flux, and pure heavy jets
as these predict too weak non-thermal X-ray emission.  Although the
pair number density is larger than the proton number density, the jets
would be dynamically dominated by the protons.

Circular and linear polarisation observations of jets can be used to
constrain the low-energy particle distribution, the magnetic field
strength and the particle content. Depending on whether this is
intrinsic to the synchrotron emission or produced by Faraday
conversion of linear polarisation to circular, different limits can be
set on the low energy particle distribution (e.g. \citealt{Homan05}).

\citet{Wardle98} measured circular polarisation from 3C~279 and, considering that
most likely this results from Faraday conversion, set an upper
limit  $\gamma_{\rm min}\lesssim20$, which would
be evidence for electron-positron jets in this source. Further
observations of PKS~0528+134, 3C~273 and even 3C~84, detected circular
polarisation, but no limits on $\gamma_{\rm min}$ have been determined
\citep{Homan99, Homan04}.  In any case, more recently 
\citet{Ruszkowski_02} showed that the observations of 3C~279 could be
consistent with both types of jet, by arguing that the linear and
circular polarisations observed could, depending on the field
configuration, be consistent with different plasma compositions. Therefore, no strong conclusion can
(yet) be drawn on such measurements.

While we have no robust explanation on the discrepancy between 
results supporting electron-proton dominated jets and our findings, one intriguing 
possibility is that jets associated with powerful sources are energetically dominated by a
proton component, while low power radio sources, such as M87 and
Perseus, are predominantly composed of an electron-positron pair
plasma.

\subsection{Magnetic Field Energy}

In general, it is possible that jets are magnetically dominated, 
with negligible contribution to the
energetics from particles (e.g. \citealt{Blandford02}). \citet{Sikora05} argued 
instead that conversion from Poynting-flux dominated to matter
dominated jets takes place very close to the black hole.

From arguments of energy equipartition between the
emitting leptons and the magnetic field, $B_{\rm eq}\sim10^{-6}$ to
$10^{-3}\G$ are estimated depending on the jet scale.  Studying M87,
\citet{Stawarz05} find a lower limit $B>3\times10^{-4}\G$ in the
brightest knot of the jet at $\sim2\kpc$ from the core, concluding that it is likely that this
represents a departure from equipartition, such that the knot is at
least marginally magnetically dominated.  

Any Poynting flux would not be easily detectable from the
radiative properties of the jet (except possibly from a population
of particles with a narrow energy distribution, as expected following acceleration from reconnection events)
and because of that no conclusion
could be drawn from the current data.

In this work we have assumed that the magnetic
field does not dominate the energy content.

\section{Particle content of radio lobes}\label{radio_particle}

The results imply that the jets in M87 and 3C~84 are light
(electron-positron), at least at their base. \citet{Celotti02} quantify the particle content of the radio bubbles in
the following way.  By comparing the pressure inside and outside of
the bubble, any extra particles beyond those inferred from the radio
emission can be determined.  This method gives estimates on $k/f$,
where $f$ is the volume filling factor of the relativistic
plasma and $k$ accounts for the extra particles.  \citet{DunnFabian04,
DunnFabian05} find $1\lesssim k/f\lesssim 1000$ (for M87 $k/f\sim10$ and
$k/f\sim500-1000$ for Perseus) and that
the radio lobes in these clusters are not in equipartition but are
particle dominated.

If the jets start as a pure electron-positron plasma, then for protons
to be present in the lobes, some entrainment of material has to occur.
This is likely to be a stochastic process and highly dependent on the
environment surrounding the jet within the inner few $\kpc$.  Powerful
sources are unlikely to pick up much material as their jets would
clear out the regions surrounding them and weaker sources would be
expected to pick up more material.  Entrainment may be the key
reason why the latter sources appear as FR~I, as indeed detailed modelling
indicates (e.g. \citealt{Laing02}).  In principle, entrainment could
slow down a jet to sub-relativistic speeds.  The amount of
entrainment, however, especially on large scales where the jets are no
longer fully collimated, is uncertain.  The exact shape of the low-energy lepton
spectrum will also affect $k/f$, as it will the matter content of
the jet.  

\citet{DeYoung06} suggest that the fact that $k>>1$ (for $f=1$) in some
clusters results from a population of cold protons present in the
jet.  This is different to the jet matter content suggested here.
This solution does have problems as the accompanying relativistic
electrons may cause jet to
decollimate unless magnetic confinement occurs.  Another solution \citet{DeYoung06}
suggest are Poynting-flux dominated jets, though these also have
associated difficulties, including whether they can produce the
 FR~I morphologies observed (e.g. \citealt{Komissarov_99,Leismann_05}).

The radio lobes of M87, Cygnus A and Fornax A appear very filamentary and
so the filling factor of the radio emitting plasma may vary, which may
have bearing on the matter content of the jet. The net effect is not
obvious as it would depend on the actual structure of the relativistic
component and its possible confinement. While clearly the kinetic
luminosity and synchrotron emission in the optically thin regime only
depends on the total particle number, the synchrotron self-absorption
estimates are also affected by the density inhomogeneities along the
line of sight.

\section{Conclusions \& Outlook}\label{concls}

If the conditions observed at the base of the jet are typical over the
timescale for bubble creation, and all the caveats discussed in the
previous section hold, then the jets in M87 and Perseus are likely to
be dominated by an
electron-positron plasma.  Other radio sources in
clusters analysed are too distant to afford the resolution necessary
to measure the synchrotron flux from the base of the jet.

Despite the significant potential offered by estimates of the jet
power from its interaction with the ICM, this method is currently
limited to determining the matter content of extra-galactic radio jets
in sources in sufficiently nearby clusters of galaxies.  A concerted
effort in monitoring the cores of nearby AGN at a variety of
frequencies will help in constraining the jet matter content.  Clear
identification of the cores, measurements of the jet velocity and
limits on the variability will aid in improving the parameters and
assumptions involved in these calculations.
Future perspectives include high accuracy measurement of resolved 
radio cores via the space VLBI, and higher frequencies observations (though the flux 
should not be dominated by optically thin emission). 

A definitive result on the matter content of jets remains, however,
elusive for now.

\section*{Acknowledgements}

We thank the anonymous referee for comments which greatly improved
this manuscript and Mary Erlund and Yuri Kovalev for stimulating discussions.  ACF,
RJHD and AC acknowledge support from The Royal Society, PPARC and the
Italian MIUR, respectively. This research was supported in part by the National Science 
Foundation under Grant No. PHY99-07949; AC acknowledges the KITP (Santa Barbara) for kind hospitality.

\bibliographystyle{mnras}
\bibliography{/home/rjhd2/bibtex/mn-jour,/home/rjhd2/bibtex/dunn}

\end{document}